\documentclass[11pt,twoside]{article}
\usepackage[pdftex]{graphicx}
\usepackage{amsmath}
\usepackage{amssymb}
\usepackage{cite}
\usepackage{lineno}

\begin{document}
\newcommand{\pst}{\hspace*{1.5em}}

\newcommand{\rigmark}{\em Journal of Russian Laser Research}
\newcommand{\lemark}{\em Volume 30, Number 5, 2009}

\newcommand{\be}{\begin{equation}}
\newcommand{\ee}{\end{equation}}
\newcommand{\bm}{\boldmath}
\newcommand{\ds}{\displaystyle}
\newcommand{\bea}{\begin{eqnarray}}
\newcommand{\eea}{\end{eqnarray}}
\newcommand{\ba}{\begin{array}}
\newcommand{\ea}{\end{array}}
\newcommand{\arcsinh}{\mathop{\rm arcsinh}\nolimits}
\newcommand{\arctanh}{\mathop{\rm arctanh}\nolimits}
\newcommand{\bc}{\begin{center}}
\newcommand{\ec}{\end{center}}

\thispagestyle{plain}

\label{sh}


\begin{center} {\Large \bf
\begin{tabular}{c}
NONRECIPROCAL
\\[-1mm]
 AND TEMPERATURE-TUNABLE LIGHT ABSORPTION
 \\[-1mm]
 IN ALAS/ITO/GAAS HYBRID METASURFACES
 
\end{tabular}
 } \end{center}

\bigskip

\begin{center} {\bf
Yi Chen$^{1}$, Lin Cheng$^{1,*,\dagger}$, Kun Huang$^{1}$, Jingyin Li$^{1}$,  Jinmin Li$^{1,2,3,4}$, 
}\end{center}

\medskip

\begin{center}
{\it
$^1$State Key Laboratory of Dynamic Testing Technology, North University of China, \\Taiyuan, China, 030051

\smallskip

$^2$Center of Materials Science and Optoelectronics Engineering, University of Chinese Academy of Sciences, \\
Beijing 100049, China

\smallskip
$^3$Research and Development Center for Wide Bandgap Semiconductors, Institute of Semiconductors, Chinese Academy of Sciences,\\ Beijing 100083, China
\smallskip

$^4$State Key Laboratory of Dynamic Measurement Technology, North University of China, \\Taiyuan 030006, China
\smallskip

$^*$Corresponding author: e-mail:kiki.cheng@nuc.edu.cn\\$\dagger$ These authors contributed equally to this work.}
\end{center}

\begin{abstract}\noindent
The single-band high-efficiency light absorption of
nanostructures finds extensive applications in various
fields such as photothermal conversion, optical sens-
ing, and biomedicine. In this paper, a vertically stacked
nanohybrid structure is designed with aluminum ar-
senide (AlAs), indium tin oxide (ITO) and gallium ar-
senide (GaAs) stacked, and the photon absorption char-
acteristics of this structure under near-infrared light
at a single wavelength of 1240 nm are exploredbased
on the finite difference time domain (FDTD) method.
When AlAs, ITO, and GaAs are stacked and incident
light enters from the GaAs side, a local light enhance-
ment phenomenon occurs. The absorption rate can
reach 91.67\%, and the temperature change rate reaches
55. 53\%, allowing for a wide-range regulation of the
structure’s absorption rate by temperature. In addition,
the AlAs/ITO/GaAs sandwich-type hybrid structure also
exhibits obvious nonreciprocity. With the change in tem-
perature, the absorption rate of different structural sizes
varies differently. The structure can be optimized and
designed according to the requirements, providing new
ideas for the design of multifunctional optoelectronic
devices.
\end{abstract}

\medskip

\noindent{\bf Keywords:}
ITO/AlAs/GaAs hybrid metasurfaces; Nonreciprocal light absorption; Temperature tunable; Finite - difference time - domain method; Light absorption characteristics; Nanohybrid structure

\pst
High-efficiency single-band light absorption has shown extremely important application values in many cutting-edge fields, such as photothermal conversion~\cite{1}, biomedical imaging, and materials science ~\cite{2}, and also plays a crucial role in aspects such as wide-field interferometric photothermal microscopy ~\cite{3}.  Taking single-walled carbon nanotubes as an example, they have extremely high optical absorption rates in the near-infrared band. This characteristic enables them to significantly enhance the photothermal killing effect in tumor cells~\cite{4}, which has strongly promoted the development of related medical treatment technologies.

Traditional single-band absorbers mainly achieve light absorption by relying on the localized surface plasmon resonance (LSPR)~\cite{5} of noble metal nanoparticles (such as Au and Ag), or by means of photonic crystal structures.   However, these traditional absorbers have many insurmountable drawbacks.  Integrating dynamic tunability and high optical transparency into a single microwave absorber has been an extremely crucial challenge up to now~\cite{6} and their absorption bandwidths are extremely narrow and the absorption rate cannot be adjusted , limiting the effective utilization of light with different wavelengths~\cite{7}. In order to break through these limitations of static devices, researchers have embarked on a new exploration direction.  By designing hybrid material systems, for example, combining semiconductors, dielectrics and metals to construct asymmetric nanostructures, and skillfully integrating external stimuli such as temperature~\cite{8} and electric fields~\cite{9}, dynamic regulation of light absorption can be achieved. Among them, temperature, as a non-invasive regulation method, has attracted much attention.  It can affect the light absorption properties of materials by changing the carrier concentration or dielectric constant of the materials ~\cite{10}. 

At present, most of the existing non-reciprocal light absorption structures, such as unidirectional absorbers, rely on magneto-optical materials or nonlinear optical effects to achieve their functions~\cite{11}.  However, these solutions have obvious drawbacks.  They often require strong magnetic fields or high-power laser excitation, which severely limits the integrated application of devices. In some cases, excessive energy input can also cause damage to the materials, which severely affects the nonreciprocal light absorption performance of the devices and reduces their stability and reliability ~\cite{12}. Therefore, how to successfully achieve temperature-regulated non-reciprocal light absorption through the collaborative design of materials and structures under low-energy-consumption conditions has become a key challenge that urgently needs to be overcome in this field. 

\begin{figure}[htb!]
\centering\setlength{\belowcaptionskip}{-5pt}  
{\includegraphics[width=0.8\linewidth]{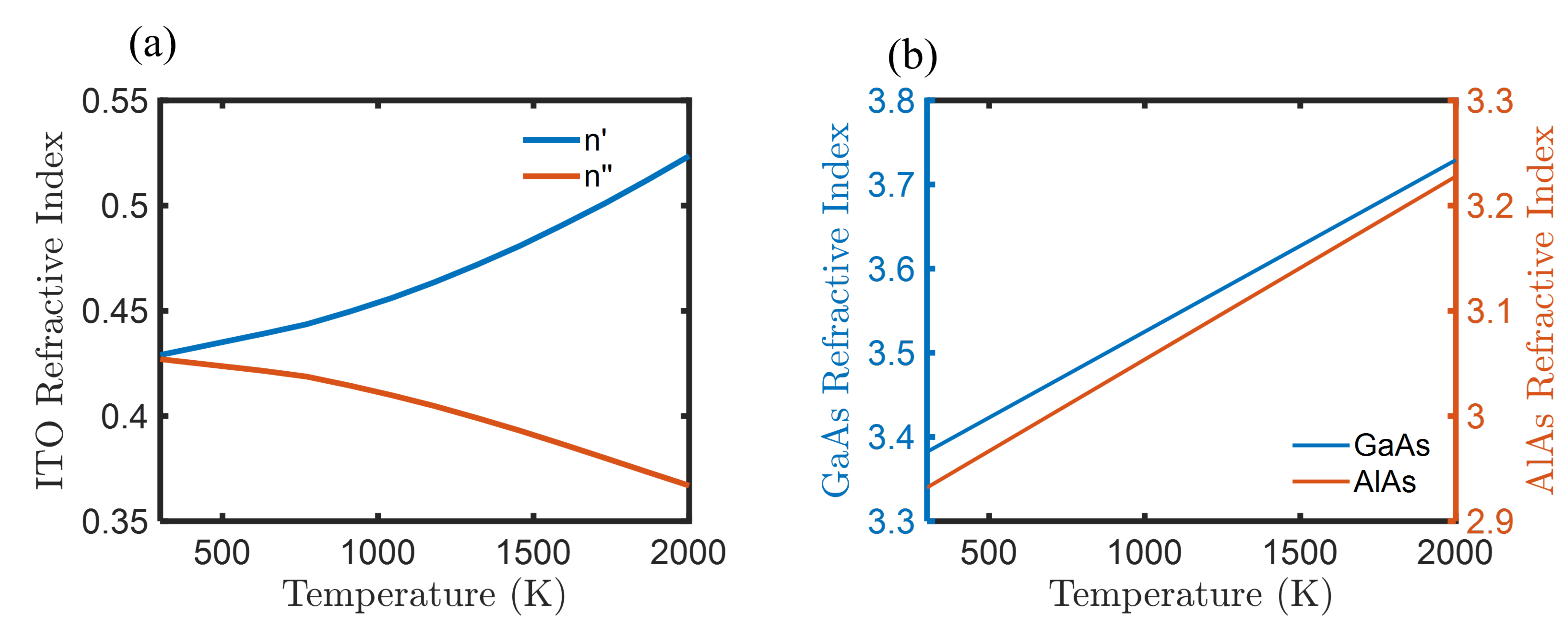}}
\caption { (a) Graph of the refractive index of ITO as a function of temperature~\cite{13}. (b) Graphs of the refractive indices of AlAs and GaAs as a function of temperature~\cite{14}.}
\label{Fig.1}
\end{figure}

\begin{figure}[htb!]
\centering
\setlength{\belowcaptionskip}{-5pt}  
{\includegraphics[width=0.8\linewidth]{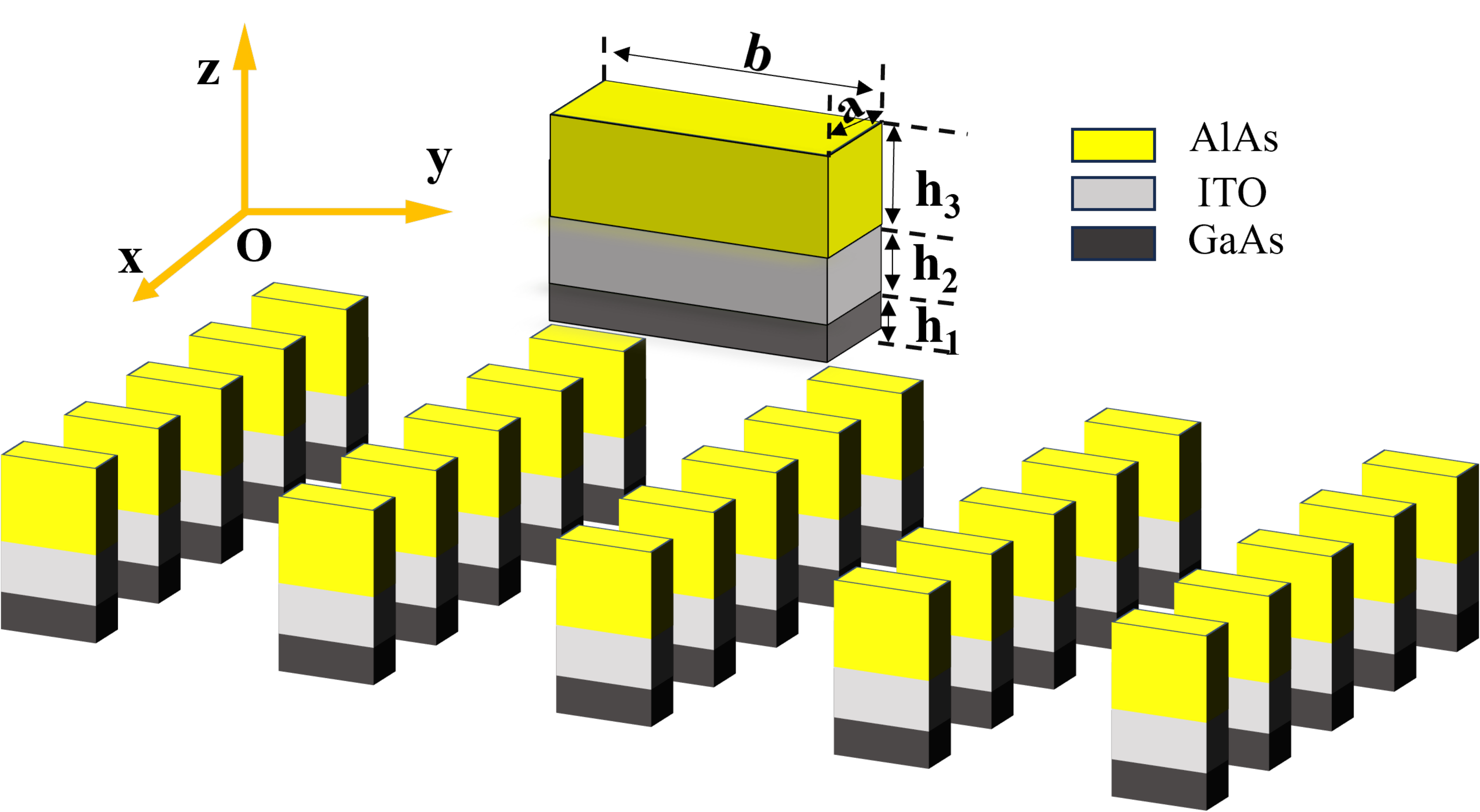}}
\caption{Schematic diagram of the vertically stacked nano-hybrid structure of AlAs/ITO/GaAs. a = b = 600 nm, $h_1$ = $h_3$ = 150 nm, $h_2$ = 600 nm.}
\label{Fig.2}
\end{figure}

As an outstanding transparent conductive material, Indium Tin Oxide (ITO) has extremely high electrical conductivity and excellent optical transparency, as shown in Figure \ref{Fig.1}a ~\cite{15}.  This unique property enables it to efficiently transport carriers.  The perfect balance between its optical transmittance and electrical conductivity provides indispensable support for the development of modern electronic technology and various optoelectronic devices~\cite{16}.  GaAs as an important semiconductor material, can exhibit higher light absorption efficiency when formed into a nano-array, thanks to the combined effects of intrinsic anti-reflection and effective excitation of resonance modes~\cite{17}. As a \text{III} - \text{V} group compound semiconductor material, the light absorption ability of AlAs is significantly enhanced after forming a heterostructure ~\cite{18}. The refractive indices of GaAs and AlAs are shown in Figure \ref{Fig.1}b. Therefore, this paper innovatively proposes a nano-hybrid structure based on vertical stacking of AlAs/ITO/GaAs, as shown in Figure \ref{Fig.2}. This structure skillfully combines the energy-band characteristics of semiconductors (AlAs, GaAs) with those of ITO.  Through precise adjustment of the thickness of the ITO layer and the stacking order, dynamic regulation of light absorption efficiency by temperature is successfully achieved.  Further research reveals that this sandwich - like structure exhibits significant non - reciprocity when the temperature changes.  That is, the difference in the absorption rates between forward and backward incident light can be dynamically adjusted with a change in the ITO size. This unique characteristic opens up new ideas for the development of non-reciprocal optoelectronic devices that do not require an external magnetic field and can be dynamically reconfigured, providing new insights for the design of multifunctional optoelectronic devices.

The FDTD algorithm is a numerical algorithm based on the time domain, which is used to solve electromagnetic field problems.  processes the time-domain differential equations in Maxwell's equations by approximating them with a second-order accurate central difference. In this way, the differential equations that originally needed to be solved directly are transformed into relatively easy-to-handle difference equations, and these difference equations are solved iteratively~\cite{19}.  We adopted the FDTD algorithm to obtain the optical properties of hybrid structures in three different structural forms, namely AlAs/ITO, ITO/GaAs, and AlAs/ITO/GaAs. Meanwhile, we obtain the transmittance (T) and reflectance (R). The absorptance (A) is then calculated using Poynting's law (A = 1 - T - R), and the influence of two factors, temperature and thickness of ITO, on the photon absorption characteristics is discussed.

\begin{figure}[htb!]
\centering
\setlength{\belowcaptionskip}{-5pt}  
{\includegraphics[width=0.8\linewidth]{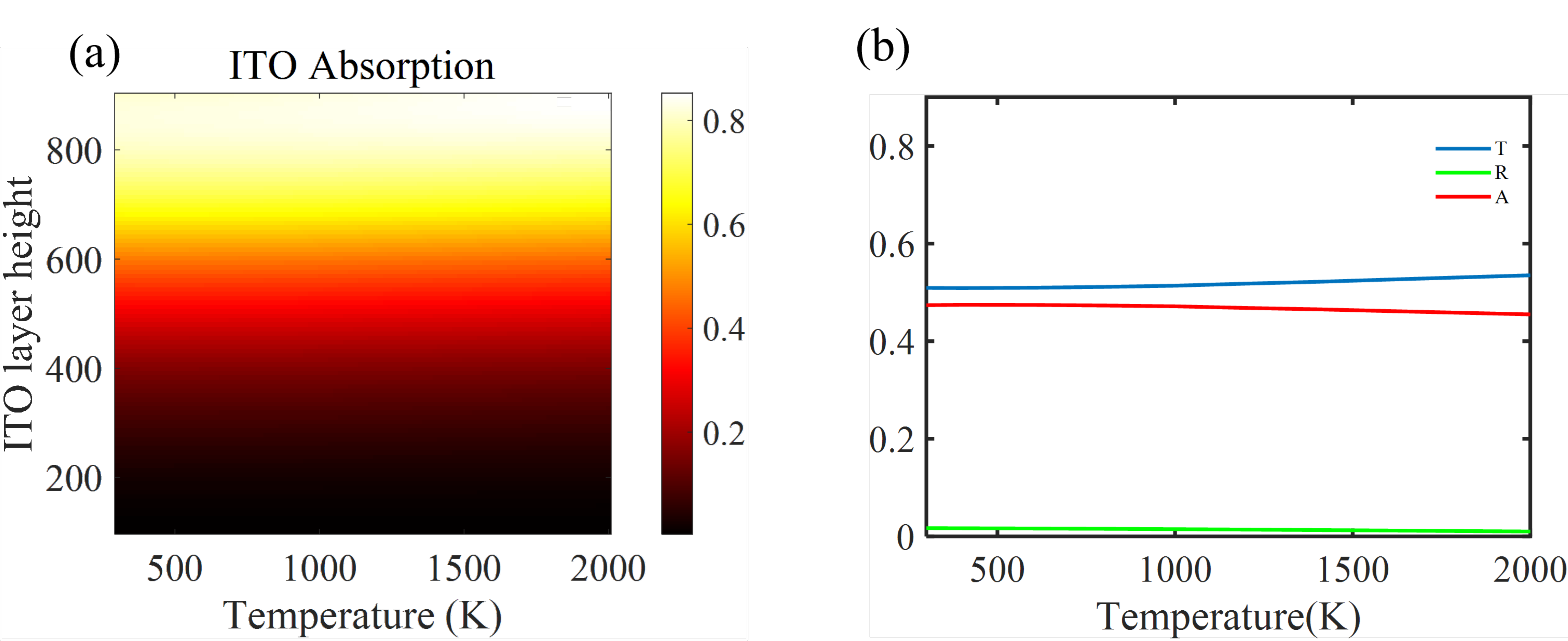}}
\caption{(a) Heatmap of the absorption of a cubic ITO with side length L ranging from 100 to 900 nm as a function of temperature. (b) Graphs of transmittance (T), reflectance (R), and absorptance (A) as a function of temperature when L = 600 nm.}
\label{Fig.3}
\end{figure}

In order to obtain the appropriate size of ITO, we explored the relationship among the side length of the cubic ITO structure, temperature, and absorptance.  The results are shown in Figure \ref{Fig.3}a: The absorptance of ITO does not change significantly with a change in temperature, and it keeps increasing as L increases.  Based on Mie scattering theory ~\cite{20}, under the light of 1240 nm, the ITO size parameter of 600 nm makes the scattering efficiency high, and the ratio of the scattering cross-section to the absorption cross-section is appropriate. The distribution of its scattered light is conducive to the re-entry of light into the material.  Multiple internal scattering prolongs the light propagation path and enhances the interaction, synergistically improving the absorption efficiency of the light at 1240 nm.  Therefore, we select ITO with L = 600 nm to establish the hybrid structure.

\begin{figure}[htb!]
\centering
\setlength{\belowcaptionskip}{-5pt}  
{\includegraphics[width=0.8\linewidth]{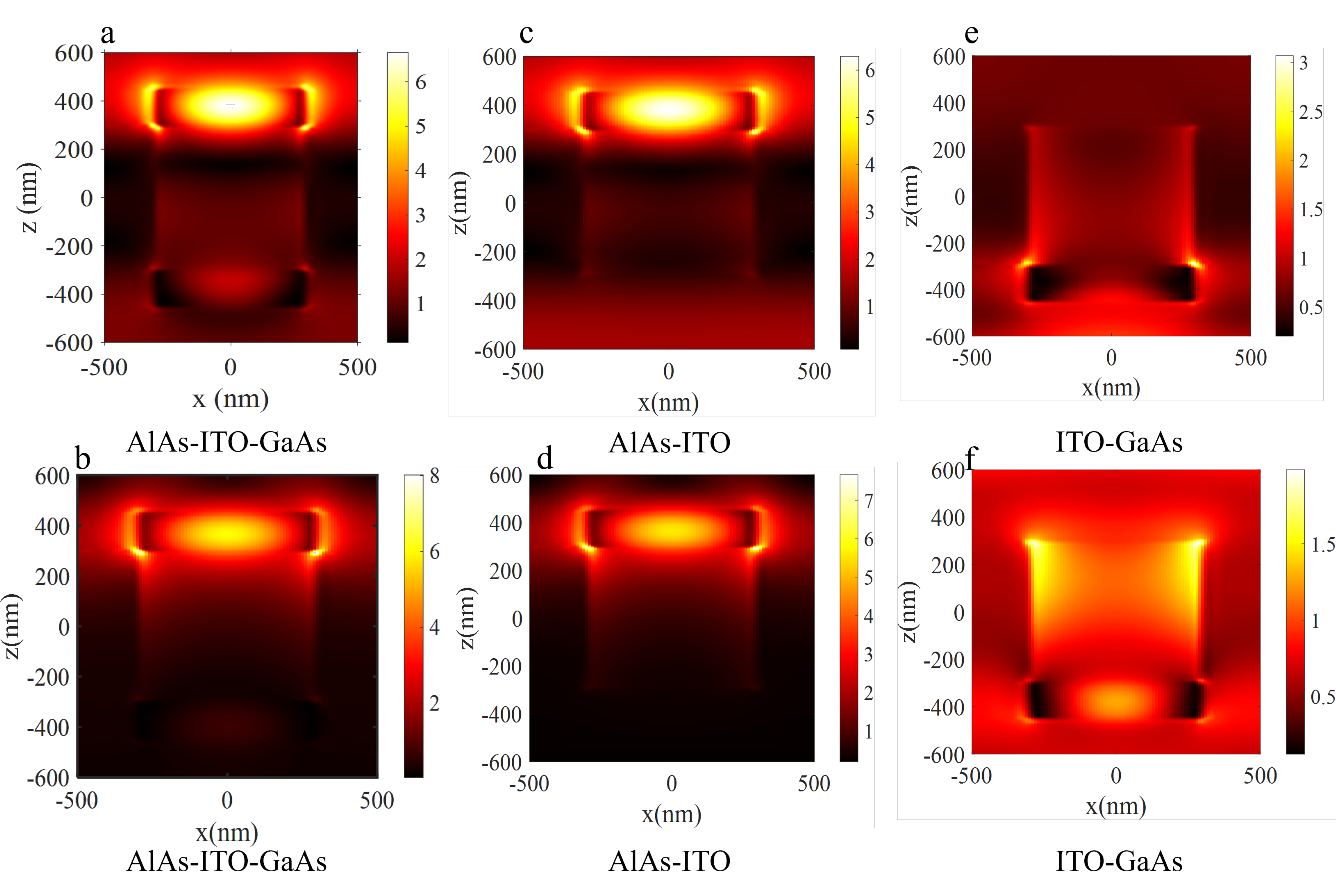}}  
\caption{At 300K and 0 polarization, the electric field distribution maps of three different structures are shown. The first row is for the light source on the GaAs side, and the second row is for the light source on the AlAs side.}
\label{Fig.4}
\end{figure}

\begin{figure}[htb!]
\centering
\setlength{\belowcaptionskip}{-5pt}  
{\includegraphics[width=0.8\linewidth]{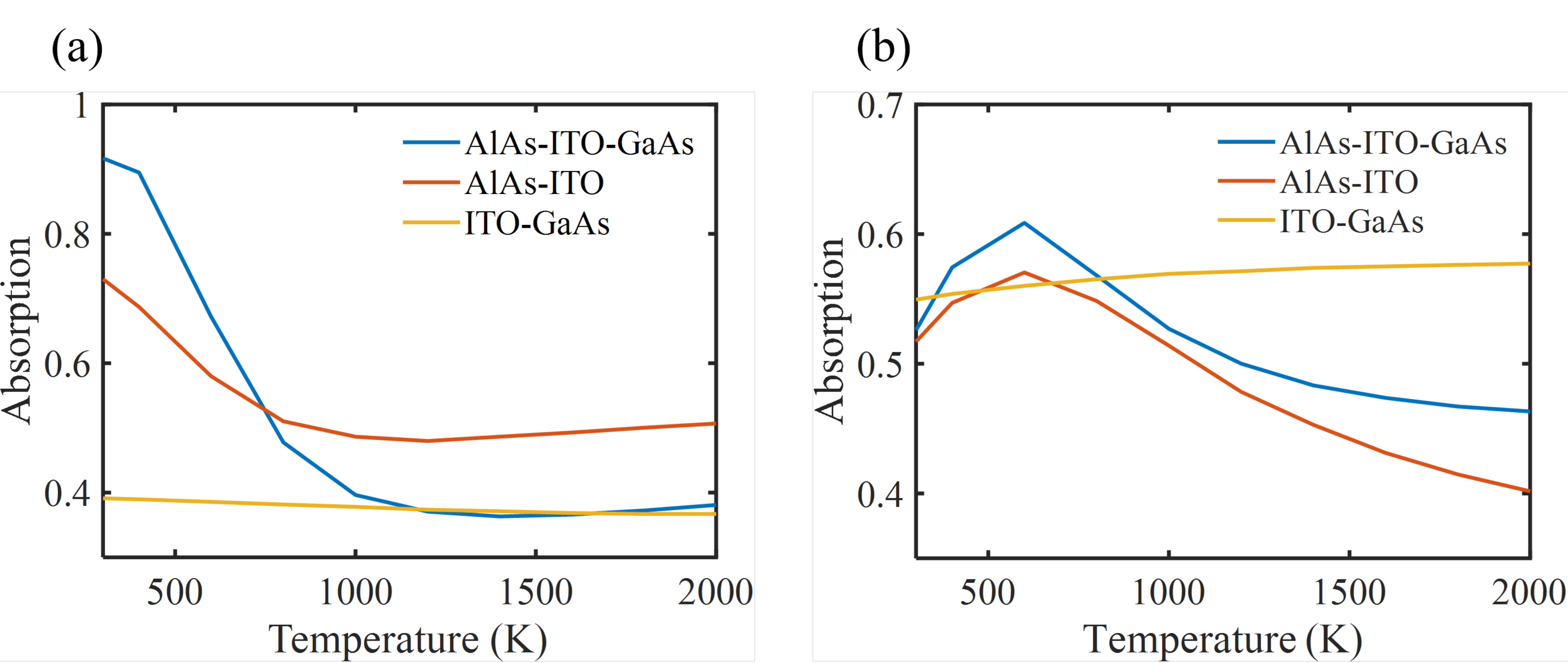}}
\caption{(a) Graph of the absorption (A) of three structures as a function of temperature when the light source is incident from the GaAs side. (b) Graph of the absorption (A) of three structures as a function of temperature when the light source is incident from the AlAs side.}
\label{Fig.5}
\end{figure}

\begin{figure}[htb!]
\centering
\setlength{\belowcaptionskip}{-5pt}  
{\includegraphics[width=0.8\linewidth]{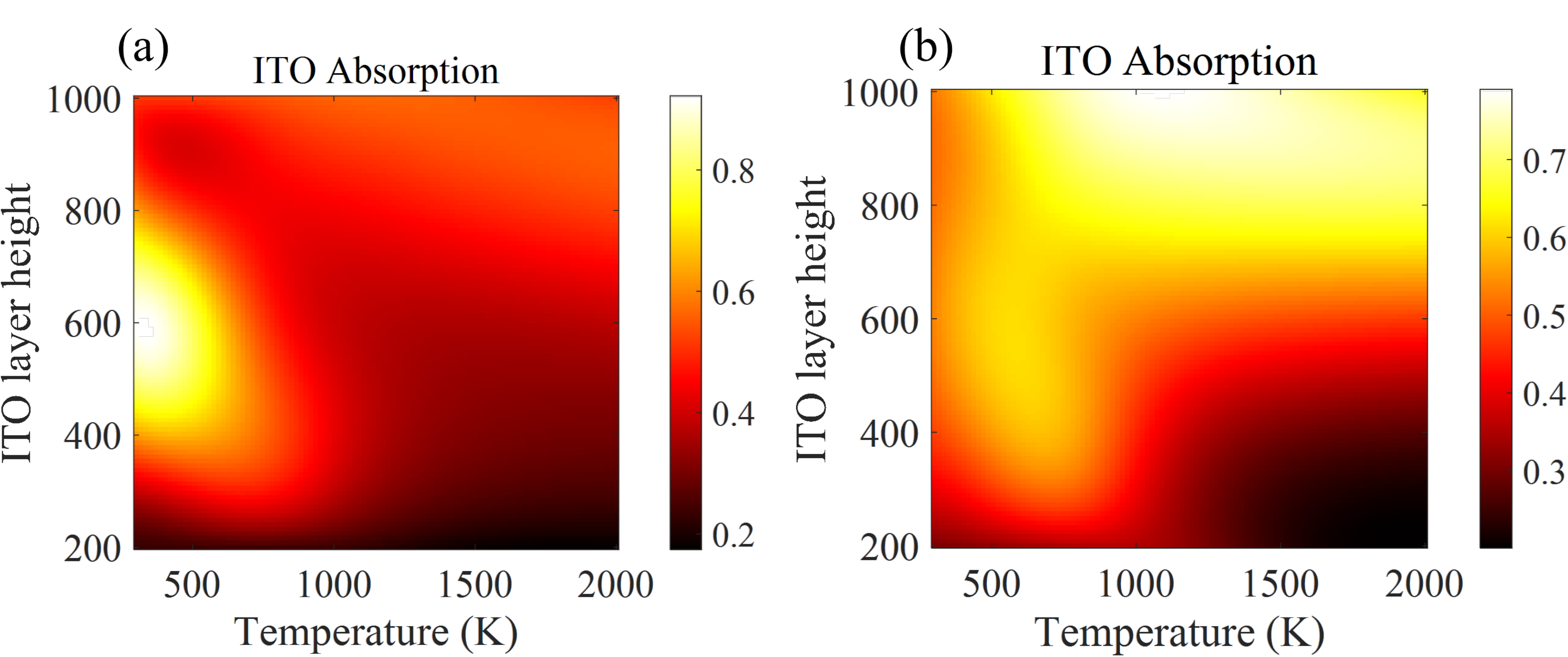}}
\caption{ Heatmap of the absorption and temperature of the AlAs/ITO/GaAs structure with the ITO thickness ranging from 100 to 900 nm. (a) The light source is on the GaAs side. (b) The light source is on the AlAs side. }
\label{Fig.6}
\end{figure}

Figure\ref{Fig.5} shows the absorption rate of the three structures varying with temperature. If the light passes through the semiconductor thin film first,  the light field is localized within the semiconductor layers on both sides,The Figure demonstrates\ref{Fig.4}.  The surface plasmon polaritons (SPPs) excited in ITO are not fully excited, and the absorptance is relatively low (the curves of ITO-GaAs in Figure \ref{Fig.5}a and AlAs-ITO in Figure \ref{Fig.5}b).when light passes through a thick ITO layer first,the surface plasmon polariton (SPP) excited by ITO couples with the photon mode of GaAs. forming hybrid plasmon-photon polaritons and the absorption effect increases significantly (the curves of AlAs-ITO in Figure\ref{Fig.5}a)and ITO-GaAs in Figure \ref{Fig.5}b)). The enhancement of the light field in the hybrid mode can improve the light absorption and transmission efficiency~\cite{21,22}. In the GaAs/ITO structure, when the light source was on the GaAs side (Figure \ref{Fig.4}e), the electric field intensity peak was higher, but the high-intensity area was smaller.  Conversely, when the light source was on the ITO side(Figure \ref{Fig.4}f), the electric field intensity peak was lower, but the high-intensity area was larger.  This extensive high-intensity distribution significantly enhanced the optical field intensity in the hybrid mode, thereby improving the light absorption efficiency and promoting a 21\% increase in the absorption rate.  Due to the Varshni effect, the band gap of GaAs redshifts with increasing temperature ~\cite{23}, and the photon energy of 1240 nm gradually approaches the bandgap, so the absorptance rises.  However, at the same time, the carrier scattering intensifies at high temperatures, reducing the overall absorption efficiency~\cite{24}. These two factors are in a balanced state, so the absorption of this combination is not sensitive to temperature changes, being less than 2\%.  It can be applied to optical detection scenarios with high requirements for temperature stability ~\cite{25}.  In the AlAs/ITO combination, the electric field distribution is analogous to that in the GaAs/ITO structure (Figure \ref{Fig.4}c d).  When the light source is on the ITO side, the absorption rate is significantly increased.  As the temperature rises, the absorption rate peak can be enhanced by 16\%.  Due to the relatively high carrier concentration of AlAs, the carrier scattering intensifies when the temperature rises~\cite{26}, resulting in a decrease in absorptance.  The absorption is sensitive to temperature changes, and the maximum absorptance varying with temperature can reach 25.02\%. The absorption can be regulated by temperature.  In the AlAs/ITO/GaAs structure, placing GaAs on the incident light side results in localized field enhancement.  There are noticeable electric fields in the media on both sides (Figure \ref{Fig.4}a b), making the hybrid structure's absorption more evident and reaching an absorption rate of 91.67\% (the AlAs-ITO-GaAs curve in Figure \ref{Fig.5} a).  Its high absorptance can play an important role in the field of solar cells~\cite{27,28,29}. The rate of change with temperature can reach 55.53\%, and the regulatory effect of temperature on absorptssion is further enhanced, thus achieving a large-scale regulation of absorption by temperature.  later we exploredthe thickness of ITO from 200 to 1000 nm, and the absorptance results extracted are shown in (Figure \ref{Fig.6}).  The structure exhibits significant non - reciprocity.  With the change of temperature, the absorptance changes of different structural sizes are different.  By matching the temperature change, the absorption peak can be found, and the structure can be optimized and designed according to the requirements.  

In summary, this paper designs a nanohybrid structure composed of vertically stacked AlAs, ITO, and GaAs and obtains its photon absorption characteristics under near-infrared light at 1240 nm.  The results of the three different structures show that when light passes through the thin semiconductor film because of the high refractive indices of GaAs and AlAs, the light field is localized within the semiconductor layers on both sides and the surface plasmon polaritons (SPPs) excited in ITO are not sufficient, resulting in a low absorption rate.  
 When passed through the thick ITO, the excited SPPs in the ITO couple with the photon modes of the GaAs to form a hybrid mode, and the absorption increases significantly.  Moreover, the enhancement of the light field improves the light absorption and transmission efficiency.  Among them, the absorption rate of the GaAs/ITO structure increases by 21\%. Because of the balance between the red shift of the GaAs bandgap and the carrier scattering at high temperatures, the absorption changes slightly with temperature.  The maximum value of the absorption rate of the AlAs/ITO combination increases by 16\%, but due to the high carrier concentration that leads to intensified scattering at high temperatures, the absorption changes significantly with temperature.  When the three are stacked and the incident light is on the GaAs side, the local light is enhanced, the absorption rate can reach 91.67\%, and the temperature change rate can reach 55.53\%, realizing a wide-range regulation of the structure's absorption rate by temperature.  When the thickness of ITO is simulated from 200 nm to 1000 nm, obvious non-reciprocity is presented, and the structure can also be optimized and designed according to requirements, providing new ideas for the design of multifunctional optoelectronic devices.  
\textbf{Ethical Approval}:not applicable

\textbf{Confilicts of Interest}: The authors declare no confilct of interest.

\textbf{Declarations}
 The statements, opinions and data contained in all publications are solely those of the individual
author(s) and contributor(s) and not of MDPI and/or the editor(s). MDPI and/or the editor(s) disclaim responsibility for any injury to
people or property resulting from any ideas, methods, instructions, or products referred to in the content.

\section{Data Availability Statement }
Data is available on request from the authors.
\section{Funding}
We would like to acknowledge the support of the National Natural Science Foundation of China (62305312), the Shanxi Province Natural Science Foundation,ina (202203021222021), the Research Project supported by the Shanxi Scholarship Council of China (2312700048MZ), and the Doctoral Science Foundation Foundation fellowship(2022M722923). Shanxi Provincial Teaching Reform and Innovation Project (2024YB008).
\section{Acknowledgement}
{Lin Cheng thanks Rasoul Alaee for his help. }


\end{document}